\documentclass{emulateapj}
\slugcomment{{\sc Accepted to ApJ:} Jan 27, 2007}

\def\m{{\rm\,m}}

\def\gm{{\rm\,g}}

\def\AU{{\rm\, AU}}

\def\yr{{\rm\,yr}}

\begin{document}

\shortauthors{Ford \& Chiang}
\shorttitle{Ice Giants}

\title{The Formation of Ice Giants in a Packed Oligarchy:\\
Instability and Aftermath}
\author{Eric B.\ Ford\altaffilmark{1,2} and Eugene I.~Chiang\altaffilmark{3,4}}
\altaffiltext{1}{Harvard-Smithsonian Center for Astrophysics,
Mail Stop 51,
60 Garden Street,
Cambridge, MA~02138, USA}
\altaffiltext{2}{Hubble Fellow}
\altaffiltext{3}{Center for Integrative Planetary Sciences,
Astronomy Department,
University of California at Berkeley,
Berkeley, CA~94720, USA}
\altaffiltext{4}{Alfred P.~Sloan Research Fellow}

\email{eford@cfa.harvard.edu, echiang@astron.berkeley.edu}

\begin{abstract}
  As many as 5 ice giants---Neptune-mass planets composed of
  $\sim$90\% ice and rock and $\sim$10\% hydrogen---are thought to
  form at heliocentric distances of $\sim$10--25 AU on closely packed
  orbits spaced $\sim$5 Hill radii apart. Such oligarchies are
  ultimately unstable. Once the parent disk of planetesimals is
  sufficiently depleted, oligarchs perturb one another onto crossing
  orbits. We explore both the onset and the outcome of the instability
  through numerical integrations,  including dynamical friction
  cooling of planets by a planetesimal disk whose properties are
  held fixed.  To trigger instability and the ejection of the first
  ice giant in systems having an original surface density in oligarchs
  of $\Sigma \sim 1 \gm/{\rm cm}^2$, the disk surface density $\sigma$
  must fall below $\sim$$0.1\gm/{\rm cm}^2$. Ejections are
  predominantly by Jupiter and occur within $\sim$$10^7\yr$. To eject
  more than 1 oligarch requires $\sigma \lesssim 0.03 \gm/{\rm cm}^2$.
  For certain choices of $\sigma$ and initial semi-major axes of
  planets, systems starting with up to 4 oligarchs in addition to
  Jupiter and Saturn can readily yield solar-system-like outcomes in
  which 2 surviving ice giants lie inside 30 AU and have their orbits
  circularized by dynamical friction.  Our findings support the idea
  that planetary systems begin in more crowded and compact
  configurations, like those of shear-dominated oligarchies.
  In contrast to previous studies, we identify
  $\sigma \lesssim 0.1 \Sigma$ as the regime relevant for understanding the
  evolution of the outer solar system, and we encourage future studies
  to concentrate on this regime while relaxing our assumption of a fixed
  planetesimal disk.
  Whether evidence of the instability can be found in Kuiper
  belt objects (KBOs) is unclear, since in none of our simulations
  do marauding oligarchs excite as large a proportion of KBOs
  having inclinations $\gtrsim 20^{\circ}$ as is observed.

\end{abstract}

\keywords{celestial mechanics---Kuiper belt---planets and satellites :
formation---solar system : formation}

\section{INTRODUCTION}
\label{sec_intro}

Without gravitational focussing, {\it in situ} coagulation
of Uranus and Neptune takes too long to complete.
In a minimum-mass disk at heliocentric distances of 20--30 AU,
timescales to assemble the ice giants
exceed the age of the solar system by 2 orders
of magnitude, if growth is unfocussed
(e.g., Goldreich, Lithwick, \& Sari 2004, hereafter GLS04).\footnote{
The problem does not disappear by merely raising the disk mass
above the minimum-mass value, since the gravitationally unfocussed growth rate
scales only linearly with the disk surface density.}
N-body coagulation simulations that do not damp
relative velocities between planetesimals, either by dynamical friction,
inelastic collisions, or gas drag, fail to form Uranus and Neptune
(Levison \& Stewart 2001; see also Lissauer et al.~1995).
The ice giants contain 10--20\% hydrogen by mass, a fraction
so large that such gas must originate from the solar nebula.
The outer planets must therefore form within a few $\times$ $10^7\yr$, before
all of the nebular hydrogen photoevaporates
(Shu, Johnstone, \& Hollenbach 1993; Matsuyama, Johnstone, \& Hartmann 2003).

One way to alleviate (but not necessarily eliminate)
the timescale problem is to form Uranus and Neptune
closer to the Sun, where material densities and collision rates are greater.
Thommes, Levison, \& Duncan~(1999, 2002) explore a scenario
in which the two planets form at distances of 5--10 AU,
between the cores of Jupiter and Saturn. Once the gas giant cores
amass their envelopes, they scatter the ice giants outward onto
eccentric orbits. These orbits subsequently circularize by
dynamical friction with planetesimals at 15--30 AU. Tsiganis et al.~(2005)
propose an alternative history in which Uranus and Neptune accrete
at 12 and 17 AU, are thrown outward by Jupiter and Saturn, and have
their orbits circularized by dynamical friction. According to their story,
the outward scattering of ice giants is triggered by having
Jupiter and Saturn divergently migrate across their mutual 2:1 resonance.

Another approach to solving the timescale problem is to consider
how gravitational focussing can be amplified. GLS04 adopt this route
by appealing to a massive disk of sub-km-sized planetesimals, similar to those
produced by coagulation simulations set in the outer
solar system (Kenyon \& Luu 1999). The disk envisioned by GLS04 has a mass
several times the minimum-mass value in condensates
so that the ``isolation mass''---the mass
to which a protoplanet grows by consuming all material within its annulus
of influence---equals Neptune's mass. The small bodies comprising
the disk collide so frequently that their velocity dispersion
damps to values less than the Hill velocity of a Neptune-mass planet.
Accretion rates then enjoy maximal enhancement by gravitational focussing,
and proto-Neptune can accrete the last half of its mass in $\sim$$10^5\yr$
(see eqn.~105 of GLS04). Gas drag supplies another means to damp
planetesimal velocity dispersions (Rafikov 2004; Chambers 2006; see also
appendix A of GLS04).

Strongly focussed, {\it in situ} assembly of planets
from a dynamically cold disk
carries, however, a potential embarrassment of riches: The disk
can spawn more ice giants than the solar system's
current allotment of 2 (Uranus and Neptune). We estimate that
about 5 isolation masses
or ``oligarchs,'' each having the mass of Neptune, can form between 15
and 25 AU (see eqn.~\ref{eqn_define_a} below).
These planets comprise a ``shear-dominated oligarchy,''
so-called because the encounter velocities between
planets and planetesimals are given by their minimum values
set by Keplerian shear. Initially, the oligarchs'
nested orbits would be stabilized by dynamical friction with the disk.
GLS04 suggest that excess oligarchs would be purged from the outer solar
system by an eventual dynamical instability. According to their
order-of-magnitude analysis, this ``velocity instability''
occurs once the mass of
the disk becomes less than the mass in oligarchs, whereupon dynamical
friction ceases to stabilize the system against mutual gravitational
stirring (a.k.a. ``viscous stirring''). In the ensuing chaos,
several oligarchs would be ejected, either by other oligarchs or by Jupiter
or Saturn, possibly leaving two survivors whose orbits could
circularize by dynamical friction at 15--30 AU.

Despite their disparate perspectives on the timescale problem and different
motivations, the scenarios of Thommes et al.~(1999), Tsiganis et al.~(2005),
and GLS04 share quite a few features. In their simplest forms,
each theory starts with a more crowded
configuration for solar system planets than is observed today; each is
characterized by an intermediate period of dynamical chaos during
which Uranus and Neptune execute highly eccentric orbits; and
each invokes final regularization of ice giant orbits by dynamical
friction with an ambient disk. Thommes et al.~(2002) and Chiang
et al.~(2006, hereafter C06) point out that these violent
histories might be encoded in Kuiper belt objects (KBOs). In particular,
so-called scattered KBOs possess large eccentricities,
inclinations, and perihelion distances which might
reflect gravitational stirring by marauding ice giants.

The notion that planets originate in compact and crowded configurations
is bolstered by the study of extra-solar systems as well.
To explain the striking preponderance of large orbital
eccentricities observed among extra-solar giant planets
(Butler et al.~2006),
multiple planets each having on the order of a Jupiter mass are imagined
to have once resided on orbits sufficiently close that the planets
scatter one another onto elliptical trajectories
(Marzari \& Weidenschilling 2002;
Ford, Rasio, \& Yu 2003; Ford, Lystad, \& Rasio 2005).

GLS04 outline a possible formation history for Uranus and Neptune in a
packed oligarchy,
and C06 expand upon its consequences for the Kuiper belt,
by making many simplifying assumptions and order-of-magnitude approximations.
In this paper, we test some of their ideas by
numerical simulations. In particular, we seek answers to the following
questions:

\begin{enumerate}
\item For a shear-dominated oligarchy containing more than two
Neptune-mass oligarchs
beyond Saturn's orbit, what is the critical value of the disk surface density
below which the velocity instability occurs?
\item What is the likelihood that the instability will result in
the survival of two oligarchs whose final orbits resemble those of Uranus
and Neptune?
\item To what degree is the Kuiper belt dynamically excited by
velocity-unstable
oligarchs?
\end{enumerate}

Our methods are described and tested in \S\ref{sec_method}. That
section contains empirical determinations
of how rapidly 5 oligarchs viscously stir one another,
with and without the gas giants Jupiter and Saturn. Comparisons
are made with analytic theory. Results of hundreds
of simulations designed to provide statistical answers to
the above questions are presented
in \S\ref{sec_result}. We summarize and offer an outlook
in \S\ref{sec_sum}.

\section{METHOD AND TESTS}
\label{sec_method}

To guide the reader,
we provide a condensed description of our method in \S\ref{sec_overview}.
Details are elaborated upon in \S\S\ref{sec_visc_stir}--\ref{sec_dyn_fric}.

\subsection{Overview}
\label{sec_overview}

We simulate the final stages of oligarchy by numerically integrating
the trajectories of 5 closely packed Neptune-mass oligarchs, together with
those of 2 gas giants resembling
Jupiter and Saturn. Oligarchs and gas giants are referred to
as planets.  We employ the hybrid orbit integrator MERCURY6 (Chambers
1999), which combines a conventional Bulirsch-Stoer
integrator to handle close encounters between planets, with
the fast symplectic algorithm
invented by Wisdom \& Holman (1991).
Viscous stirring of an oligarch by other planets is simulated
as accurately as the orbit integrator solves the gravitational
equations of motion. Case studies of viscous stirring
are described in \S\ref{sec_visc_stir}.

To model dynamical friction between a planet and the surrounding
disk of planetesimals, we introduce a perturbative force on each
planet. The force damps the component of the planet's velocity that
differs from the local disk (circular) velocity. For simplicity, we take the
disk to have a constant surface density between
an inner and an outer radius.
Disk parameters are held
fixed. In our simple scheme,
the planets respond to the disk through dynamical
friction, but the disk does not
respond to the planets. The details of the
perturbation force are provided in \S\ref{sec_dyn_fric}.
The validity of our fixed disk approach is briefly considered
in \S\ref{sec_reduce}.

Finally, to investigate how oligarchs might excite the Kuiper belt, an
ensemble of test particles is included in a subset of
the simulations. We use the
terms ``test particle'' and ``Kuiper belt object (KBO)''
interchangeably. These test particles are intended to represent large
KBOs like those observed today, having sizes on the order of 100 km. This
size is small enough that we can neglect dynamical friction between
KBOs and the disk, yet also large enough that we can ignore damping of
KBOs' velocities by physical collisions with the disk (C06). Thus, in
our simulations, KBOs (test particles)
feel directly only the gravity of the Sun and of the planets.

In the sub-sections below,
we explore separately the processes of
viscous stirring (\S\ref{sec_visc_stir}) and dynamical friction
(\S\ref{sec_dyn_fric}), in isolation from one another.
We present full-fledged simulations, in which the two
processes are combined, in \S\ref{sec_result}.

\subsection{Viscous Stirring}
\label{sec_visc_stir}

We first study how multiple
oligarchs gravitationally stir one another.
For this sub-section, we ignore dynamical friction with the disk, but include
the gas giants, Jupiter and Saturn.
We compare our findings to those of GLS04. The results of this
sub-section will be applied in \S\ref{sec_result} to understanding
the threshold conditions required
for velocity instability.

\subsubsection{Initial Conditions}
\label{sec_init_con}

We consider $N_{\rm olig} = 5$
oligarchs, each having the mass of Neptune ($\mu = m/m_{\odot} = m_{\rm
N}/m_{\odot} = 5.1 \times 10^{-5}$). The oligarchs
are initially spaced 5 Hill radii apart in semi-major
axis ($a$); that is, the difference between semi-major axes of
nearest-neighboring oligarchs is

\begin{equation}
a_{j+1} - a_j = 2.5 (\mu/3)^{1/3}a_j + 2.5 (\mu/3)^{1/3}a_{j+1} \,,
\label{eqn_define_a}
\end{equation}

\noindent where $j$ ranges from 1 to $N_{\rm olig}$.
The coefficient of 2.5 is
inspired by numerical studies by Greenberg et al.~(1991) of the
width of a protoplanet's feeding zone, for the case where
Keplerian shear dominates the relative velocity between a
protoplanet and a planetesimal (see their eqn.~9;
see also Ida \& Makino 1993).
While the coefficient of 2.5
is the standard value for our study, the coefficient in reality
can be somewhat different, depending on the accretion and migration
histories of the planets. We explore some consequences of varying
the coefficient in \S\ref{sec_stir_curve} and \S\ref{sec_instab_ss}.

We assume the semi-major axis of the innermost oligarch
$a_1 = 15$ AU. Then according to eqn.~(\ref{eqn_define_a}),
semi-major axes of the next four oligarchs equal 17.1,
19.4,
22.1,
and 25.1
AU.  Initial eccentricities, and initial inclinations relative to an arbitrary
reference (x-y) plane, are such that $e_{j} = \sin i_{j} = 10^{-4}$.
All orbital elements in this paper are osculating and
referred to a barycentric coordinate system. That is,
when computing the osculating Kepler elements for a given body,
the position and velocity of the body are referred to the system barycenter
(calculated using all massive bodies),
while the central mass of the assumed Kepler potential
equals the mass of the Sun plus that of the given body (alone).
For each oligarch,
the initial longitude of ascending node, longitude of pericenter, and
mean anomaly are randomly generated from uniform distributions between
0 and $2\pi$.

Jupiter and Saturn are assigned initial masses and semi-major axes
equal to their current values: $\mu_{\rm J} = m_{\rm J}/m_{\odot} = 9.5
\times 10^{-4}$, $\mu_{\rm S} = m_{\rm S}/m_{\odot} = 2.9 \times
10^{-4}$, $a_{\rm J}= 5.18$ AU, and $a_{\rm S} = 9.54$ AU.
Initial eccentricities equal $e_{\rm J}=e_{\rm S} =0.05$, and
inclinations relative to the reference plane are such that $\sin
i_{\rm J} = \sin i_{\rm S} = 0.01$. Orbital longitudes are randomly generated,
just as they are for the oligarchs.

These initial conditions, particularly our choices for $a_1 = 15 \AU$ and
$N_{\rm olig} = 5$, are somewhat arbitrary. They are intended to represent
qualitatively the final stages of shear-dominated oligarchy in the outer
solar system. We will adjust starting parameters
(e.g., $a_1$, $N_{\rm olig}$) in later sections to achieve simulation
outcomes in better agreement with observed
properties of the solar system.

A total of $N_{\rm real} = 200$ orbital integrations
(``realizations'') are performed with the hybrid integrator MERCURY6
(Chambers 1999), each characterized by a unique set of starting
longitudes and each lasting $10^7 \yr$.  The timestep for the
symplectic integrator is set to 130 days. Timesteps for the
conventional Bulirsch-Stoer integrator are as short as necessary to
achieve an accuracy parameter of $10^{-10}$.  The changeover distance
that separates the symplectic regime from the close encounter regime
is set to $\Delta r_{\rm crit} = 3$ Hill radii.

A ``collision'' between two
bodies occurs when their mutual separation becomes less than the sum
of their physical radii. The physical radius of each oligarch is
computed using Neptune's bulk density of $1.6 \gm/{\rm cm}^{3}$. Physical
radii for Jupiter and Saturn are computed using densities
of $1.3$ and $0.7 \gm/{\rm cm}^{3}$, respectively.  We assume that bodies that
collide merge completely.

An ``ejection'' occurs when an oligarch's distance from the
Sun exceeds 10000 AU and when its total kinetic plus potential energy
(evaluated in barycentric coordinates with the potential energy set
to zero at infinity) becomes positive. Ejected planets are dropped
from the simulation.

\subsubsection{Results: Outcomes After $t = 10^7 \yr$}
\label{sec_fates}

At $t = 10^7\yr$, the outcomes for all
$N_{\rm all} = N_{\rm real} \times N_{\rm olig} = 200 \times 5 = 1000$
oligarchs divide into the following mutually exclusive categories,
in order of decreasing frequency of incidence:

\begin{enumerate}
\item Ejection but no collision (463)
\item No collision and no ejection (439)
\item Collision with another oligarch but no ejection
(42; i.e., 21 oligarchs remain, each with mass twice
that of an original oligarch)
\item Collision with Sun only (19)
\item Collision with Jupiter only (14)
\item Collision with another oligarch, and subsequent ejection
or subsequent collision with Jupiter or Saturn (13)
\item Collision with Saturn only (10)

\end{enumerate}

The dominant outcome is ejection. In 50\% of the realizations
(i.e., 100 out of $N_{\rm real}=200$), at least one oligarch
is ejected before $t=1.6 \times 10^6\yr$.
By $t = 3.2 \times 10^6 \yr$, 85\% of all realizations
experience a first ejection. All but 5 out of 200 realizations
experience at least one ejection of an oligarch by $t = 10^7\yr$.

Jupiter and Saturn are responsible for the preponderance of
ejections. When we repeat the experiment with Jupiter and Saturn
omitted, outcomes at $t = 10^7 \yr$ are as follows: 870
out of 1000 oligarchs experience neither an ejection nor a collision, 118
collide with another oligarch (so that 59 remain), and 12 collide with
two other oligarchs (so that 4 remain).  No ejection is observed to
occur by $t = 10^7 \yr$ when the gas giants are absent.

These outcomes are consistent with timescale estimates by GLS04.
Neglecting Jupiter and Saturn,
GLS04 predict that the oligarch ejection timescale is $\sim$$10^9\yr$ (see
their
eqn.~114). This is consistent with our finding that no ejection occurs
within $t = 10^7\yr$ in the absence of gas giants. GLS04 mention
the possibility that Jupiter and Saturn hasten ejections.
We confirm this possibility. When gas giants are present, we find
the ejection timescale is $\sim$$10^6\yr$.

\subsubsection{Results: Eccentricity and Inclination Growth (``Stirring
Curves'')}
\label{sec_stir_curve}

\begin{figure}
\epsscale{0.9}
%\plotone{StirTest_final2.eps}
\plotone{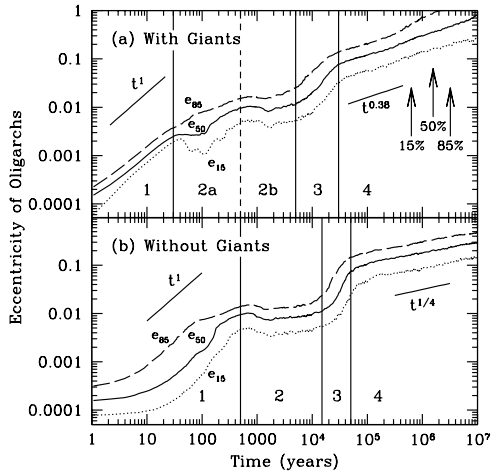}
\caption{Viscous stirring of oligarchs, in the absence of dynamical friction,
for an oligarch separation of $5R_{\rm H}$.
We show the median (solid curve), 85$^{\rm th}$ percentile
(long dashed curve), and 15$^{\rm th}$ percentile (dotted curve)
of the oligarchs' eccentricity distribution versus time, based on N-body
integrations with no dynamical friction.
In panel (a), each integration includes Jupiter, Saturn, and five oligarchs.
In panel (b), only the five oligarchs are integrated. Arrows in panel (a)
mark the times when 15\%, 50\%, and 85\% of $N_{\rm real} = 200$
realizations (integrations) experience the first ejection of an oligarch.
Vertical lines separate various phases of evolution discussed
in \S\ref{sec_stir_curve}.
}
\label{fig_stir}
\end{figure}

Fig.~\ref{fig_stir} tracks the median eccentricity,
$e_{50}(t)$, of all oligarchs.  The sample from which the median is
drawn always contains $N_{\rm all} = 1000$ objects, regardless of
whether oligarchs collide or are ejected. When computing the median,
we adopt the following rules: ejected oligarchs have their
eccentricities set equal to 1 (but remain part of the sample); an
oligarch that collides with either the Sun, Jupiter, or Saturn has its
eccentricity set equal to 1; and oligarchs that collide with other
oligarchs are still counted as separate objects and are each assigned an
eccentricity equal to the current eccentricity of the merged body.
Experiments with alternative sets of rules produced no qualitative
changes to our results.

\placefigure{fig_stir}

The resultant ``stirring curves'' of Fig.~\ref{fig_stir} exhibit a variety
of behaviors.
We first discuss the case when Jupiter and Saturn are omitted from the
integrations (Fig.~\ref{fig_stir}b). As annotated in Fig.~\ref{fig_stir}b,
we distinguish four phases of viscous stirring:

\begin{enumerate}
\item {\it Distant Conjunctions:} At early times $t \lesssim 500 \yr$,
  planetary orbits do not cross and $e_{50}$ grows roughly linearly
  with time. A linear dependence is expected from viscous stirring by
  distant conjunctions, i.e., conjunctions between oligarchs that are
  not nearest neighbors. To derive the $t^1$ scaling, we estimate using
  the impulse approximation that a conjunction between two oligarchs
  separated by distance $x < a$ imparts eccentricities on the order of
  \begin{equation}
    \Delta e \sim \mu \left( \frac{a}{x} \right)^2
  \end{equation}
  to both bodies, provided they have eccentricities less than $\Delta
  e$ prior to conjunction. For a given oligarch, a total of $N
  \sim \Sigma a x / m$ oligarchs all reside about the same distance
  $x$ away, where $\Sigma$ is the mass surface density of oligarchs.
  Conjunctions with these oligarchs occur over the synodic period
  \begin{equation}
    t_{\rm syn} \sim \frac{a}{x}\, t_{\rm orb}\,,
  \end{equation}
  where $t_{\rm orb}$ is the orbital period. Then
  \begin{equation}
    \frac{de}{dt} \sim N \frac{\Delta e}{t_{\rm syn}} \sim \frac{\Sigma
a^2}{m_{\odot}} \frac{1}{t_{\rm orb}} \sim {\rm constant}
    \label{eqn_distant}
  \end{equation}
  as roughly observed in Fig.~\ref{fig_stir}.
  As time elapses, ever closer neighbors at smaller $x$
  drive the stirring. This reasoning matches that given by
  GLS04 in their treatment of viscous stirring in the shear-dominated
  regime; their eqn.~(33) is identical in form to our eqn.~(\ref{eqn_distant}).
  The $t^1$ scaling is also derived by Collins \& Sari (2006)
  and Collins, Schlichting, \& Sari (2007). These latter studies concentrate
  on the limit $N_{\rm olig}\gg 1$.

\item {\it Conjunctions with Nearest Neighbors:} At intermediate times
  $500 \lesssim t ({\rm yr}) \lesssim 1.5\times 10^4$, planetary orbits remain
  non-crossing but the eccentricity distribution hardly changes. Since
  the synodic period between nearest neighboring oligarchs is $t_{\rm
    syn} \sim 500\yr$, a given oligarch during this phase
  experiences repeated conjunctions with its nearest neighbor.  Such
  repeated close encounters might be expected to produce chaotic
  motion and to cause eccentricities to random walk, in which case
  $e_{50} \propto t^{1/2}$.  That this scaling is not observed
  implies that our 5 oligarchs
  do not behave in a strongly chaotic manner despite their
  close spacing. Indeed, we observe in our simulations that the epicyclic
phases
  (true anomalies) of a given oligarch at successive
  conjunctions with its nearest
  neighbor do not vary completely randomly.
  Perturbations from conjunctions with a nearest neighbor
  apparently tend to cancel out during
  this second phase.

\item {\it Onset of Orbit Crossing:} The cancellations characterizing
  the preceding
  phase are not perfect, however. Eventually, from $t \sim
  1.5\times 10^4\yr$ to $t\sim 5 \times 10^4\yr$, eccentricities surge as
  oligarchs start crossing orbits.  The median eccentricity $e_{50}$
  surpasses the orbit-crossing value, $e \approx 0.06$, during this
  third phase.

  Our finding that orbits cross in a few $\times$ $10^4\yr$ is consistent
  with numerical experiments by Chambers, Wetherill, \& Boss (1996),
  who measure times required for close encounters to occur in initially
  circular, co-planar, multi-planet systems as a function of planet mass
  and orbital spacing. For reference, our spacing of $5 R_{\rm H}$ corresponds
  to 4 ``mutual Hill sphere radii'' as defined by those authors.

\item {\it Orbit Crossing:} At late times $t \gtrsim 5 \times 10^4
  \yr$, oligarchs routinely cross orbits and we observe $e_{50}
  \propto t^{0.25}$.  We can reproduce this scaling using the following
  particle-in-a-box argument. An oligarch's random velocity $v$ at
  time $t$ is determined largely by its closest encounter with another
  oligarch up until that time. We call the impact parameter
  characterizing this closest encounter $b_{\rm min}$. From
  kinetic theory, $n b_{\rm min}^2 v t \sim 1$, where $n \sim \Sigma
  \Omega / m v$ is the number density of oligarchs and we have assumed
  that the random velocity distribution of oligarchs is
  isotropic.  It follows that $b_{\rm min} \propto 1/t^{1/2}$ and
  $v \sim (Gm/b_{\rm min}^2)^{1/2} \propto t^{1/4}$. This scaling
  agrees with that of eqn.~(49) of GLS04 and
  that of eqn.~(14) of C06.

\end{enumerate}

When we restore Jupiter and Saturn to the integrations (Fig.~\ref{fig_stir}a),
we can still discern the four phases enumerated above. However, compared to the
case without giants, some phase boundaries are shifted to earlier
times, and $e_{50}$ rises more quickly during some phases. Phase 1
transitions to phase 2a at $t \approx 30 \yr$; at this time, all
oligarchs have undergone their first conjunctions with Jupiter and
Saturn. Phase 2a transitions to phase 2b at $t \approx 500 \yr$; as in the
case without giants, this transition marks the time when every oligarch
has experienced about one conjunction with its nearest neighboring
oligarch. During phase 2b, we witness the same remarkable near-constancy
of $e_{50} \approx 0.01$. Finally,
during phase 4 at $t \gtrsim 3 \times 10^4 \yr$, when oligarchs
are on crossing orbits, $e_{50} \propto
t^{0.38}$. Such growth outpaces that observed in the absence of
the gas giants.

What about oligarch inclinations?
In simulations without gas giants,
we observe that the median inclination $i_{\rm 50}$ remains fairly constant at
the initial value of $10^{-4}$ until phase 3. As orbits cross,
$\sin i_{50}$ surges up to $\sim$$0.05$, and thereafter grows as $t^{0.25}$
during phase 4, just as $e_{50}$ does.
By $t= 10^7\yr$, $i_{50} \approx 10^{\circ}$.
When Jupiter and Saturn are included, $\sin i_{50} \propto t^{0.28}$ during
phase 4. By $t = 10^7\yr$, $i_{50} \approx 10^{\circ}$, as was the case
without the gas giants.
The modest growth of oligarchs' inclinations
will limit the degree to which inclinations of KBOs
are stirred (\S\ref{sec_stir}).

\begin{figure}
\epsscale{0.9}
%\plotone{.eps}
\plotone{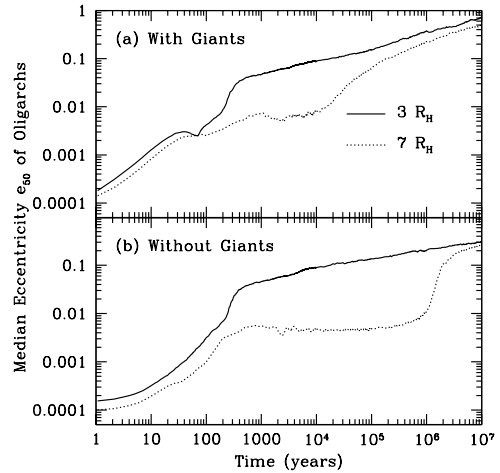}
\caption{Viscous stirring of oligarchs, in the absence of dynamical friction,
for assumed oligarchic spacings of $3R_{\rm H}$ and $7R_{\rm H}$.
Data are generated and presented in the same way as for Fig.~\ref{fig_stir},
but only median eccentricities are displayed.
Without the gas giants, the same 4 phases characterizing viscous
stirring for $5 R_{\rm H}$ (Fig.~\ref{fig_stir}) are evident
for $7 R_{\rm H}$. A spacing of $3R_{\rm H}$ produces
systems sufficiently chaotic that phase 2b is indiscernible.
% 1 (distant conjunctions)
%transitions to phase 4 (orbit crossing).
Adding the gas
giants, however, reduces the sensitivity of the results to the
oligarchic spacing.
}
\label{fig_stir2}
\end{figure}

Not all of the different phases of viscous stirring that we observe are
anticipated from the study of GLS04, which documents only the $t^1$
scaling characterizing shear-dominated oligarchy (phase 1) and the
$t^{1/4}$ scaling characterizing the super-Hill, orbit-crossing regime
(phase 4, no giants). Their analysis
misses the intermediate phase 2 of slow-to-no growth just prior to orbit
crossing, and the significant roles that Jupiter and Saturn play in
accelerating viscous stirring during phase 4 ($t^{0.38}$ vs.~$t^{1/4}$).
That differences exist is not too surprising, as their analysis is rooted
in the large $N_{\rm olig} \gg 1$ limit, whereas for our system,
$N_{\rm olig} = 5$. More importantly, we take nearest-neighboring
oligarchs to be separated by 5 Hill sphere radii,
as dictated by the extent of an oligarch's feeding zone in a shear-dominated
disk (Greenberg et al.~1991), whereas the order-of-magnitude
equations of GLS04 governing shear-dominated oligarchy assume
the separation is closer to $\sim$1 Hill sphere radius.
In this regard, we present in Figure \ref{fig_stir2} viscous stirring curves
for cases where the oligarch separation is $3 R_{\rm H}$ and $7 R_{\rm H}$
(corresponding to coefficients in eqn.~(\ref{eqn_define_a}) of
1.5 and 3.5, respectively). Without gas giants, for a separation
of $3R_{\rm H}$, phases 2 and 3 disappear, leaving only phases
1 and 4 as originally envisioned by GLS04. The time to orbit crossing
varies from $\sim$$300 \yr$ to $\sim$$2 \times 10^6\yr$ as the spacing
changes from 3 to 7 Hill radii. This extreme sensitivity
to spacing was also found by Chambers et al. (1996).
Including gas giants, however, reduces this sensitivity,
as Fig.~\ref{fig_stir2}a shows.

\placefigure{fig_stir2}

The actual oligarchic spacing might only be determined by
careful numerical simulations of accretion and orbital migration.
We adopt in this paper a standard value of $5R_{\rm H}$,
identical to that assumed by GLS04, and motivated
by studies of shear-dominated accretion by Greenberg et al.~(1991).
Shorter spacings seem less attractive insofar as
they will produce smaller isolation masses for a given disk surface density.

The intermediate phase of slow-to-no growth of eccentricity
that characterizes oligarchic spacings $\geq 5 R_{\rm H}$
will prove important in determining the threshold disk surface density
below which dynamical friction cooling cannot balance viscous heating,
i.e., the threshold surface density for the velocity instability
(\S\ref{sec_require}).

\subsection{Dynamical Friction}
\label{sec_dyn_fric}

Oligarchs grow from a disk of planetesimals. Those planetesimals that
are not accreted exert dynamical friction on oligarchs. The conditions
for velocity instability, and the ease with which survivors of the
instability return to low-eccentricity, low-inclination orbits, depend
on the strength of dynamical friction. We describe how we implement
dynamical friction in our simulations in \S\ref{sec_prescrip},
present a test case in \S\ref{sec_testcase},
and show that our implementation is compatible with the formulae
of GLS04 in \S\ref{sec_connect_gls}.

\subsubsection{Prescription}
\label{sec_prescrip}
Consider a planet having an eccentricity and an inclination much
greater than those of disk planetesimals. Dynamical friction reduces
the planet's random (peculiar) velocity: the difference $\vec{v} \equiv v
\hat{v}$ between the orbital velocity of the planet and that of the
mean disk flow. From Binney \& Tremaine (1987, their eqn.~7-17),

\begin{equation}
\frac{d\vec{v}}{dt} = - \frac{2\pi G^2 m \rho}{v^2 } \ln (1+\Lambda^2) \,
\hat{v} \,,
\label{eqn_binney_tremaine}
\end{equation}

\noindent where $\rho$ is the local mass density of the disk, $m$ is
the mass of the planet, $G$ is the gravitational constant, and

\begin{equation}
\Lambda = \frac{b_{\rm max}(v^2 + 2\langle \sin^2i_{\rm disk}\rangle
v_{\rm circ}^2)}{Gm}
\label{eqn_lambda}
\end{equation}

\noindent is the Coulomb parameter appropriate for dynamical friction in
a Keplerian disk (Stewart \& Ida 2000). For $b_{\rm max}$, the maximum
impact parameter between the planet and a disk planetesimal, we adapt
the expression of Stewart \& Ida (2000; see the discussion following
their eqn.~2-17):

\begin{equation}
b_{\rm max} = R_{\rm H} + r\left( \langle \sin^2i_{\rm disk}\rangle +
\sin^2i\right)^{1/2} \,,
\label{eqn_b_max}
\end{equation}

\noindent where $r$ is the instantaneous distance between the planet
and the system barycenter,
$R_{\rm H} = (\mu/3)^{1/3}r$ is the Hill sphere radius,
and $\langle \sin^2 i_{\rm disk} \rangle^{1/2} \ll 1$ is the
inclination dispersion of disk planetesimals, held constant for each
simulation (more on its precise value later).
The term $2\langle \sin^2 i_{\rm disk} \rangle v_{\rm circ}^2$
in eqn.~(\ref{eqn_lambda}) approximates the square of the velocity
dispersion of disk planetesimals, where
$v_{\rm circ}$ is the local mean disk speed.
We take ${v}_{\rm circ}$ to equal the speed
that the planet would have on a circular orbit about the Sun.

Usually it is assumed in writing eqn.~(\ref{eqn_binney_tremaine})
that $\Lambda \gg 1$. We do not make this assumption. In fact,
we use eqns.~(\ref{eqn_binney_tremaine})--(\ref{eqn_b_max})
regardless of the magnitude of $\Lambda$. When
$\Lambda \ll 1$, dynamical friction is in the shear-dominated regime.
In \S\ref{sec_connect_gls}, we justify our universal
application of
(\ref{eqn_binney_tremaine})--(\ref{eqn_b_max})
by showing that these equations correctly reduce
to forms appropriate to the shear-dominated case when $\Lambda \ll 1$.

We implement dynamical friction as follows.
We are interested in the case where the oligarchs are so dynamically
excited that each plunges through a vertically thin, dynamically cold
disk of planetesimals twice per orbit.
Specifically, we assume that
the time a planet spends immersed in the disk, $\Delta
t \approx h/|v_z|$, where $h$ is the full vertical thickness of the disk
and $|v_z|$ is the vertical component of $\vec{v}$
at the moment of disk crossing,
is short compared to the orbital period, $t_{\rm orb}
=2\pi/\Omega$.
Equivalently, $\sin i \gg \langle \sin^2i_{\rm
disk}\rangle^{1/2}$. At every disk crossing, a planet receives a
specific impulse of

\begin{eqnarray}
\Delta \vec{v} \,\, \approx \,\, \frac{d\vec{v}}{dt} \Delta t & \approx &
- \frac{2\pi G^2m}{v^2} \ln (1+\Lambda^2) \rho \Delta t \, \hat{v}
\nonumber \\
& \approx & - \frac{2\pi G^2 m}{v^2} \ln (1+\Lambda^2) \frac{\sigma}{|v_z|}
\,\hat{v} \,, \label{eqn_prescrip}
\end{eqnarray}

\noindent where $\sigma$ is the disk surface density
(height-integrated $\rho$).
At every timestep of the integration, we check whether the planet
crosses through the disk, which is fixed to lie in the x-y plane.
At moments of disk crossing,
we apply a kick according to eqn.~(\ref{eqn_prescrip}): we increment
the velocity of the planet by $\Delta \vec{v}$ but do not change the planet's
position. We compute the difference velocity $\vec{v}$ by subtracting
the barycentric velocity of the planet from ${v}_{\rm circ} \hat{\phi}$,
where $\hat{\phi}$ is the unit vector in the azimuthal direction.
The kick is applied in the subroutine MDT\_HY.FOR in the MERCURY6 code,
after the positions are advanced but before the velocities are updated
for the second time by the interaction Hamiltonian.

For all our simulations, we fix $\langle \sin^2i_{\rm
disk}\rangle^{1/2} = 10^{-3}$, a value sufficiently small that $\sin i
\gg \langle \sin^2 i_{\rm disk} \rangle^{1/2}$ for all but a
tiny fraction of the time.  In other words, the strength
of dynamical friction in our simulations depends much more strongly
on the planet's random speed $v$ than on the
much smaller random speeds of planetesimals
(see eqns.~\ref{eqn_lambda}--\ref{eqn_b_max}).
Planetesimals can maintain low velocity dispersions by inelastic
collisions or by gas drag.

Our scheme for dynamical friction damps orbital inclinations
relative to the x-y (disk) plane.
The inclination may become so small that
$\Delta t \propto 1 / \sin i$
exceeds $t_{\rm orb}$, at which point the planet is immersed within
the disk and our impulse approximation breaks down.
To account for this possibility, we arbitrarily set
$\Delta t = \min (h/|v_z|, 0.025/\Omega)$. Our softening prescription
slows but does not stop the damping of inclination and eccentricity
for $\sin i \lesssim 0.004$. The softening might represent
slight misalignments between the planet's orbital plane and the
disk midplane, which in reality will be warped.
We have verified that our principal findings,
described in \S\ref{sec_result}, do not depend sensitively upon
the details of this prescription.
While the precise values of the inclinations that we compute are clearly
not very meaningful, we expect that our results are still qualitatively
correct, i.e., the code correctly identifies when mutual inclinations
between planetary orbits are large $(> 1 \,{\rm rad})$ and small $(\ll 1\,{\rm
rad})$.

The main virtue of our prescription for dynamical friction is its
simplicity.  We need only specify the disk surface density $\sigma
(r)$, not the volumetric density $\rho (r,z)$, and we need only
apply dynamical friction at disk crossings. The main shortcoming of
our prescription is that it does not account for the response of the
disk to the planets.  The clearing of gaps and
the generation of time-dependent, non-axisymmetric
structures (see, e.g., Goldreich \& Tremaine 1982)
will alter the gravitational back-reaction of the disk onto
embedded planets in ways that could be significant but that we (and GLS04)
do not capture.
For ways to model the response of the
planetesimal disk more realistically, see
Lithwick \& Chiang (2007) and Levison \& Morbidelli (2007).

\subsubsection{Test with Single Planet}
\label{sec_testcase}

We test our prescription for dynamical friction in the case of a
single planet interacting with a disk. The disk has constant surface
density $\sigma = 1 {\,{\rm g}}/{{\rm cm}}^{2}$.
A Neptune-mass oligarch is placed on an orbit having
initial semi-major axis $a_{\rm init} = 20$ AU and initial
eccentricity and inclination such that $e_{\rm init} = \sin i_{\rm
init} = 0.3$. Two cases are considered, one where the initial argument
of perihelion $\omega_{\rm init} = 0$ and another where
$\omega_{\rm init} = \pi/2$. The evolution of $e(t)$ and $i(t)$
depends on $\omega_{\rm init} \bmod \pi$.

\begin{figure}
\epsscale{0.9}
%\plotone{DynFricTest_final2.eps}
\plotone{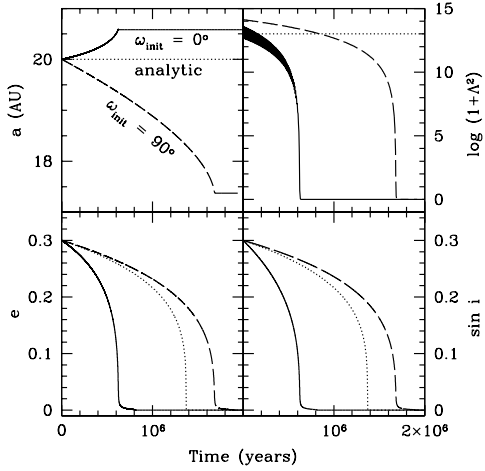}
\caption{Test of our prescription for dynamical friction. We calculate
the orbital evolution of a single Neptune-mass planet interacting with a disk
of surface density $\sigma = 1 \gm /{\rm cm}^2$. Solid curves
denote the evolution when the planet's initial argument of pericenter
$\omega_{\rm init} = 0$ (so that the orbit intersects the disk at
pericenter and apoocenter), while dashed curves correspond
to $\omega_{\rm init} = \pi/2$ (so that the orbit intersects the disk
at quadrature). Dotted curves represent our analytic solution
(\ref{eqn_analyt}) which assumes a constant semi-major axis $a$
and Coulomb parameter $\Lambda$.
}
\label{fig_DynFricTest}
\end{figure}

Fig.~\ref{fig_DynFricTest} displays the resultant evolution.
The planet's eccentricity and inclination both drop
precipitously
toward zero after a time on the order of $10^6\yr$; the exact time varies
by a factor of 3 between our choices for $\omega_{\rm init}$.
The semi-major axis can increase or decrease. It
changes by 3--13\%, on the order of but less than the
starting eccentricity.

\placefigure{fig_DynFricTest}

We check our numerical results by comparing them to the following
approximate analytic solution. Since the kick $\Delta \vec{v}$ is applied
twice per orbit, we write

\begin{eqnarray}
\frac{d\vec{v}}{dt} & \approx & - \frac{2 \Delta \vec{v}}{t_{\rm orb}}
\nonumber \\
& \approx & - 4\pi \ln(1+\Lambda^2) \frac{G^2 \sigma m}{t_{\rm orb}}
\frac{\vec{v}}{v^3 |v_z|} \,. \label{eqn_two_kicks}
\end{eqnarray}

\noindent We set $e=\sin i$ and make the following approximations:
$\vec{v} = (e\hat{p}+i\hat{z})\Omega a$, $v = \sqrt{2} e \Omega a$,
and $|v_z| = i \Omega a$. Here $\hat{p}$ and $\hat{z}$ are unit vectors
that lie parallel and perpendicular to the disk, respectively.
Then (\ref{eqn_two_kicks}) simplifies to

\begin{equation}
\frac{de}{dt} = -\frac{\ln (1+\Lambda^2)}{\sqrt{2}} \frac{G\sigma}{\Omega a}
\frac{m}{m_{\odot}} \frac{1}{e^3} \,,
\label{eqn_dedt}
\end{equation}

\noindent with an analogous equation for $i$. For fixed $\Lambda$ and $a$,
eqn.~(\ref{eqn_dedt}) integrates to

\begin{equation}
e = \left( e_{\rm init}^4 - \frac{4 \ln (1+\Lambda^2)}{\sqrt{2}}
\frac{m}{m_{\odot}} \frac{G\sigma}{\Omega a} t \right)^{1/4} \,.
\label{eqn_analyt}
\end{equation}

\noindent The eccentricity (equivalently, inclination) vanishes
in a finite time

\begin{equation}
t_{\rm vanish} = \frac{\sqrt{2}}{4\ln (1+\Lambda^2)} \frac{m_{\odot}}{m}
\frac{\Omega a}{G\sigma} e_{\rm init}^4 \,.
\label{eqn_vanish}
\end{equation}

\noindent We overlay eqn.~(\ref{eqn_analyt}) in
Fig.~\ref{fig_DynFricTest}, taking as representative values
$a = a_{\rm init} = 20$ AU and
$\ln (1+\Lambda^2) = \ln (1 + \Lambda_{\rm init}^2) = 13$. The analytic
solution
lies between the two numerical solutions. We consider
the agreement acceptable.

\subsubsection{Connecting Our Prescription to GLS04}
\label{sec_connect_gls}

Our equations can be re-cast into the same forms as those of
GLS04, under the assumption $e=i$. We start with
eqn.~(\ref{eqn_two_kicks}) and substitute $|v_z| = v/\sqrt{2}$, $t_{\rm orb}
= 2\pi/\Omega$, and $m = (4\pi/3)\rho_{\rm p} R_{\rm p}^3$, where
$\rho_{\rm p}$ and $R_{\rm p}$ are the internal density and physical
radius of the planet:

\begin{equation}
\frac{1}{v}\frac{dv}{dt} = -2^{3/2} \ln(1+\Lambda^2) \, \sigma \Omega  \,
\frac{4\pi G^2 \rho_{\rm p} R_{\rm p}^3}{3 v^4} \,.
\label{eqn_start}
\end{equation}

\noindent We next recognize that
$4\pi G^2 \rho_{\rm p}^2 R_{\rm p}^4 / 3 = 3 v_{\rm
esc,p}^4 / 16 \pi$, where $v_{\rm esc,p}$ is the escape velocity from
the surface of the planet. Then eqn.~(\ref{eqn_start}) simplifies to

\begin{equation}
\frac{1}{v} \frac{dv}{dt} = - \frac{3\sqrt{2}}{8\pi} \ln (1 + \Lambda^2)
\frac{\sigma \Omega}{\rho_{\rm p} R_{\rm p}}
\left( \frac{v_{\rm esc,p}}{v} \right)^4 \,. \label{eqn_recast}
\end{equation}

\noindent When $\ln (1+\Lambda^2)$ is a constant of order unity,
eqn.~(\ref{eqn_recast}) matches the form of eqn.~(45) of GLS04, evaluated
using the first line of their eqn.~(46),
with their planetesimal random velocity $u$ replaced by $v$
(since $v>u$; see their section 5.5, end of first paragraph). This
formula describes dynamical friction in the dispersion-dominated
regime, where $v$ exceeds the Hill velocity $v_{\rm H} = \Omega
R_{\rm H}$.

On the other hand, it is possible for $\Lambda \ll 1$. This happens,
according to (\ref{eqn_lambda})--(\ref{eqn_b_max}), when $v \sim
\sqrt{2} i \Omega \ll \sqrt{Gm/R_{\rm H}} \sim v_{\rm H}$ (terms
proportional to $\langle \sin^2 i_{\rm disk} \rangle^{1/2} = 10^{-3}$ are
negligible). In this shear-dominated regime,

\begin{equation}
\Lambda \approx \frac{R_{\rm H} v^2}{Gm} \ll 1 \,,
\end{equation}

\noindent $\ln (1+ \Lambda^2) \approx \Lambda^2$, and
eqn.~(\ref{eqn_recast}) reduces to

\begin{eqnarray}
\frac{1}{v} \frac{dv}{dt} & = & - \frac{3}{\pi \sqrt{2}}
 \frac{\sigma \Omega}{\rho_{\rm p} R_{\rm p}}
\left( \frac{R_{\rm H}}{R_{\rm p}} \right)^2 \nonumber \\
& = & - \frac{3}{\pi \sqrt{2}} \frac{\sigma \Omega}{\rho_{\rm p} R_{\rm p}}
 \frac{1}{\alpha^2} \,, \label{eqn_shear}
\end{eqnarray}

\noindent where we have defined, following GLS04, $\alpha \equiv
R_{\rm p}/R_{\rm H}$. Eqn.~(\ref{eqn_shear}) matches, to within a
numerical constant, eqn.~(45) of GLS04, evaluated using the second
line of their eqn.~(46).

We conclude that our treatment of dynamical friction
is compatible with that of GLS04.

\section{RESULTS}
\label{sec_result}

We present the results of simulations that combine viscous stirring
due to multiple oligarchs, with dynamical friction due to a planetesimal disk.

\subsection{Initial Conditions and Integration Times}
\label{sec_init}

Each system begins with either $N_{\rm olig}=5$, 4, or 3 oligarchs, together
with
the gas giants, Jupiter and Saturn. Initial conditions are the same as those
described in \S\ref{sec_init_con}, except that initial
eccentricities and inclinations of oligarchs are
such that $e_j = \sin i_j = 10^{-2}$. Initial inclinations
are therefore ten times larger
than the assumed inclination dispersion of disk planetesimals
($\langle \sin^2i_{\rm disk} \rangle^{1/2} = 10^{-3}$).
Initial eccentricities are the same as those that characterize
phase 2b of the viscous stirring curves (Fig.~\ref{fig_stir}a).
Planetary orbits initially do not cross.

Dynamical friction is exerted by a disk of constant surface density
$\sigma$ which extends from a barycentric radius of 12.5 AU to 45 AU.
The outer boundary coincides with the location of the classical
Kuiper belt (C06). The inner boundary is less well motivated. It is
chosen so that the oligarchs reside initially inside the disk while
the gas giants do not.
We explore values for $\sigma$ ranging from 0.4 to $0.001 \gm /{\rm cm}^2$.
For reference, the initial surface density in oligarchs
is $\Sigma \approx 1.5 \gm /{\rm cm}^2$.

In a subset of runs, we include 400 test particles representing
large KBOs. These feel the gravity of the planets but do not feel dynamical
friction from the disk (see \S\ref{sec_overview}).
Initial semi-major axes of test particles range
from $a = 40$ to 45 AU, and initial
eccentricities and inclinations are such that $e = \sin i = 10^{-2}$.
For all planets and KBOs, initial longitudes of ascending node,
longitudes of pericenter, and mean anomalies are randomly chosen from
uniform distributions between 0 and $2\pi$.

Settings for the MERCURY6 code are the same as those given
in \S\ref{sec_init_con}, except for the duration of integration.
The integration automatically halts when there are a catastrophic number
of ejections, i.e., when the only massive bodies remaining
include the Sun, Jupiter,
Saturn, and one oligarch (which may have collided with other oligarchs).
In the absence of such an event, each system is
integrated first to $t = 2\times 10^7\yr$. If
the planets seem to have stabilized at that time---i.e., their eccentricities
no longer grow---then we stop the integration and record
the outcome as final. Otherwise, we repeat this test as necessary
at $5 \times 10^7\yr$ and $1 \times 10^8 \yr$.

By $t = 1 \times 10^8 \yr$, most but not all realizations stabilize.
Excluding systems that are stopped abruptly once only
three planets remain, we find that $\lesssim 10\%$ of systems have
undergone a close encounter (here defined to occur when the distance
between any two planets is less than 1 Hill radius) within
the last $10^7\yr$ of the integration, for $N_{\rm olig}=5$
and all values of $\sigma$ tested. For $N_{\rm olig}=4$ and 3,
the corresponding fractions are $\lesssim 10\%$ and $\lesssim 1\%$.
Those realizations that do not stabilize by $t = 1 \times 10^8\yr$
are typically characterized by small values of
$\sigma \lesssim 0.006 \gm/{\rm cm}^2$ (i.e., relatively weak dynamical
friction) and one oligarch remaining on an eccentric orbit
that extends well past the outer edge of the disk.
For these low values of $\sigma$,
the number of ejected oligarchs that we report
will be underestimated, but not in a way that changes
our qualitative conclusions.

The initial conditions just summarized apply to all results in
the following two sections, \S\S\ref{sec_require}--\ref{sec_two}.
An alternative set of initial conditions,
motivated by the findings in those sections,
and results pertaining thereto are presented in \S\ref{sec_compact}.

\subsection{Threshold Disk Surface Densities for Instability}
\label{sec_require}

Oligarchs cross orbits when the disk surface density $\sigma$ is
so low that dynamical friction cooling cannot compete with viscous stirring.
GLS04 estimate the critical
surface density for instability to be on the order of the surface density
of oligarchs: $\sigma_{\rm crit} \sim \Sigma$ (see also Chiang \& Lithwick
2005 for a correction in the derivation of this result).
How well does this criterion predict the onset of instability for
our system of $N_{\rm olig} = 5$ oligarchs?

As noted in \S\ref{sec_visc_stir}, the rates of eccentricity growth
(viscous stirring) exhibited in our N-body integrations
differ from those estimated by GLS04. Specifically, our rates are slower,
as evidenced by the period of slow-to-no
growth of eccentricity (phase 2b) in Fig.~\ref{fig_stir}a.
Overestimating the vigor of viscous stirring leads to overestimates
for $\sigma_{\rm crit}$. We try to predict $\sigma_{\rm crit}$ ourselves
by drawing from the numerical results of \S\ref{sec_visc_stir}.
We observe in Fig.~\ref{fig_stir}a that after a time
$t_{\rm unstable} \sim 5\times 10^3 \yr$, eccentricities
surge rapidly to crossing values.\footnote{Chambers et al. (1996)
provide fitting formulae for $t_{\rm unstable}$ as functions
of oligarch mass and orbital spacing, but only for the case
without the gas giants Jupiter and Saturn.}
Therefore for oligarchs to cross orbits, eccentricities must not be allowed
to vanish by dynamical friction before
$t_{\rm unstable}$:

\begin{equation}
t_{\rm vanish} > t_{\rm unstable} \,.
\label{eqn_unstable}
\end{equation}

\noindent Using (\ref{eqn_vanish}) for $t_{\rm vanish}$,
we find that (\ref{eqn_unstable}) translates into $\sigma < \sigma_{\rm crit}$,
where

\begin{eqnarray}
\sigma_{\rm crit} & \sim & \frac{\sqrt{2}}{4\ln (1+\Lambda^2)}
\frac{m_{\odot}}{m} \frac{\Omega a e_{\rm init}^4}{G t_{\rm unstable}}
\label{eqn_semi_empirical}\\
\nonumber \\
& \sim & 0.2 \gm /{\rm cm}^2 \sim 0.1 \Sigma \,.
\label{eqn_crit}
\end{eqnarray}

\noindent The numerical evaluation takes $a = 19.4 \AU$,
$e_{\rm init} = 0.01$ (the value appropriate to phase 2b of the
viscous stirring curves in Fig.~\ref{fig_stir}a; the median
eccentricity does not rise above 0.01 for $t < t_{\rm unstable}$), and
$\ln (1+ \Lambda^2) = \ln (1 + \Lambda_{\rm init}^2) = 0.02$.

\begin{figure}
\epsscale{0.9}
%\plotone{branching_543.eps}
\plotone{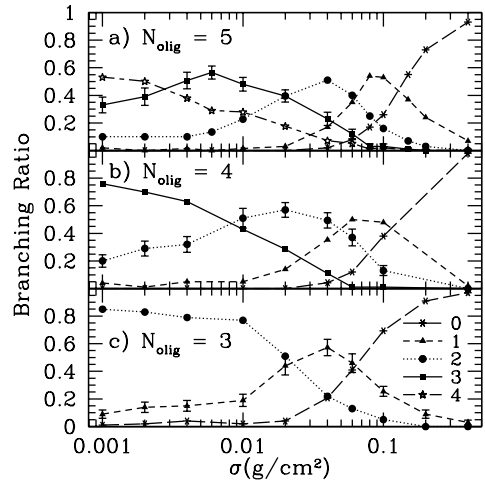}
\caption{Branching ratios for the outcome of the velocity instability,
for systems that include dynamical friction with a disk
of surface density $\sigma$. For comparison,
the initial surface density in oligarchs is $\Sigma \approx 1.5 \gm/{\rm
cm}^2$.
Systems begin with Jupiter, Saturn, and either $N_{\rm olig} = 5$ (top panel),
4 (middle panel), or 3 (bottom panel) oligarchs.
Different symbols denote the fraction of realizations
that result in 0, 1, 2, 3, or 4 ejections of oligarchs, whether
collisionally merged or not. Few oligarchs collide in these
simulations; the fraction of realizations that result in 1 collision
is less than 20\% for all values of $\sigma$ tested,
and the fraction of realizations that result in $>1$ collision
is $< 1$\%. To produce $\geq 1$ ejection with $>$ 20\% probability requires
$\sigma \lesssim \Sigma/10$. Standard error bars due to counting statistics
are given for curves corresponding to $N_{\rm olig}-2$ ejections.
}
\label{fig_branching}
\end{figure}

\begin{figure}
\epsscale{0.9}
%\plotone{RvsLogT_5_0.1_082.eps}
\plotone{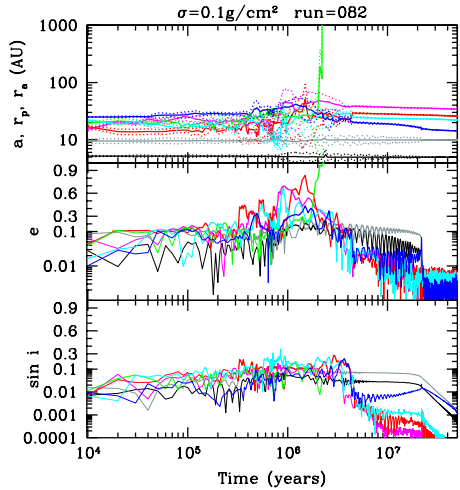}
\caption{Typical
realization for $\sigma = 0.1\gm/{\rm cm}^2 \approx 0.07 \Sigma$
in which 1 oligarch escapes before
the system stabilizes. The top panel plots semi-major axis (solid curves),
and periastron and apastron distances (dotted curves) for each of the seven
planets. Eccentricities (middle panel) and inclinations (bottom panel)
are plotted using a mixed log-linear ordinate. Black curves refer to Jupiter,
gray curves refer to Saturn, and remaining colored curves
refer to the five oligarchs.
}
\label{fig_sample_1}
\end{figure}

Our semi-empirical estimate for $\sigma_{\rm crit}$ finds support in
Fig.~\ref{fig_branching}a, which documents, for all runs starting
with $N_{\rm olig} = 5$ oligarchs, the frequency of incidence of
outcomes (``branching ratios'') as a function of $\sigma$.
Instability and the subsequent ejection of 1 and only 1 oligarch
is the dominant outcome for $\sigma = 0.1 \gm /{\rm cm}^2 \approx 0.07 \Sigma$.
For $\sigma \gtrsim 0.3\Sigma$, more than 90\% of realizations
produce no ejection. Fig.~\ref{fig_sample_1} displays
a sample simulation for $\sigma \approx 0.07\Sigma$ in which 1 oligarch
escapes before the system stabilizes.

\placefigure{fig_branching}

\placefigure{fig_sample_1}

Figs.~\ref{fig_branching}b and \ref{fig_branching}c
supply branching ratios for $N_{\rm olig} = 4$ and 3. To produce the same
number of ejections with smaller $N_{\rm olig}$ (less viscous stirring)
requires
smaller $\sigma$ (less dynamical friction). For example, for
$N_{\rm olig} = 3$, the ejection of 1 and only 1 oligarch is the dominant
outcome for $\sigma \approx 0.03\Sigma$,
occurring in about 60\% of realizations.
The corresponding $\sigma$ for $N_{\rm olig} = 4$ is $0.05\Sigma$.

\subsection{Runs with Two Surviving Oligarchs (Solar-System-Like Outcomes)}
\label{sec_two}

While $\sigma_{\rm crit} \sim 0.1 \Sigma$ roughly characterizes
the onset of instability and the subsequent ejection of a single
oligarch, the disk surface density must be reduced below $\sigma_{\rm crit}$
to produce more than 1 ejection in a large fraction of runs.
To generate an outcome reminiscent
of our solar system starting with $N_{\rm olig}=5$
requires 3 ejections and the survival of 2 oligarchs.
According to Fig.~\ref{fig_branching}a, such an outcome occurs
with a maximum probability of $\sim$50\% for
$\sigma \approx 0.01 \gm /{\rm cm}^2 \approx 0.007 \Sigma$.
The probability exceeds 20\% for all values of
$\sigma \lesssim 0.03 \Sigma$ that we tested.

Figs.~\ref{fig_branching}b and \ref{fig_branching}c
indicate that for $N_{\rm olig}=4$ and 3,
the values of $\sigma$ most likely to produce 4-planet systems (Jupiter,
Saturn, plus 2 oligarchs)
are $\sim$$0.01\Sigma$ and $\sim$$0.03\Sigma$, respectively.
The probabilities for generating 4-planet systems starting
with $N_{\rm olig} = 4$ or 3 reach large maximum values of about 50\%,
and remain above 20\% over a considerable range in $\sigma$, up
to $\sim$$0.07\Sigma$ in the case $N_{\rm olig}=3$.

The vast majority of the resultant 4-planet
systems are correctly ordered; they contain, in order of increasing
semi-major axis, Jupiter, Saturn, and 2 oligarchs. Moreover, in most of these
systems, the surviving planets have not experienced a collision.
In the following sub-sections we further quantify
the properties of these correctly ordered, collisionally unmodified,
4-planet systems, comparing them to those of the solar system.
We refer to the 2 surviving oligarchs in each of these systems
as Uranus and Neptune.

\subsubsection{Final Semi-Major Axis Distributions}
\label{sec_semi}

\begin{figure}
\epsscale{0.9}
%\plotone{FinalAvsSigma.eps}
\plotone{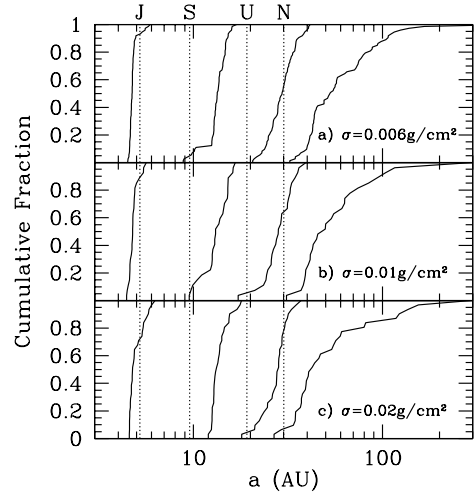}
\caption{Cumulative distributions of final semi-major axes
of systems that contain, in order of increasing semi-major axis,
Jupiter, Saturn, and two oligarchs (Uranus and Neptune),
each with their original mass.
These 4-planet systems result from integrations originally containing
7 planets ($N_{\rm olig}=5$).
Each panel corresponds to simulations performed
with the disk surface density $\sigma$ indicated.
Vertical lines mark the current semi-major axes
of the giant planets in our solar system.
Apart from a tendency for Saturn and Neptune to be pulled
toward the inner ($a = 12.5\AU$) and outer ($a = 45\AU$) edges
of the disk with increasing $\sigma$ (increasing dynamical friction),
the characteristic final semi-major axes do not depend
on $\sigma$. Saturn, Uranus, and Neptune have final orbits too large compared
to their actual counterparts in the solar system.
}
\label{fig_finalA_sigma}
\end{figure}

Because packed oligarchies evolve chaotically, we can only meaningfully compute
probability distributions for their final semi-major axes.
Fig.~\ref{fig_finalA_sigma} illustrates how these distributions depend on
$\sigma$, for realizations starting with $N_{\rm olig} = 5$.
Increasing $\sigma$ increases dynamical friction and therefore
tends to pull Saturn and Neptune, whose orbits lie
near disk boundaries, into the disk.
For example, if Saturn's orbital apocenter intersects
the disk while its pericenter remains outside the disk,
then dynamical friction will circularize the orbit
by raising the pericenter closer to apocenter.
The kinks in the distribution functions for Saturn in
Figs.~\ref{fig_finalA_sigma}a and \ref{fig_finalA_sigma}b
are located at $a=12.5$ AU,
exactly at the inner disk edge. The kink vanishes in
Fig.~\ref{fig_finalA_sigma}c. For the simulations in
Fig.\ref{fig_finalA_sigma}c,
dynamical friction is strong enough
to pull Saturn's orbit wholly into the disk at $a \geq 12.5 \AU$.

\begin{figure}
\epsscale{0.9}
%\plotone{FinalA.eps}
\plotone{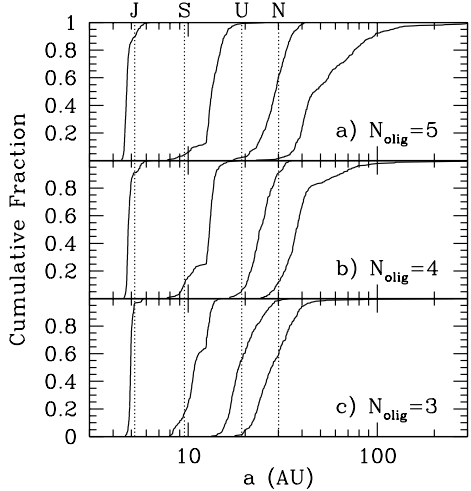}
\caption{Cumulative distributions of final semi-major axes
of correctly ordered, collision-free, 4-planet systems,
shown for different starting values of $N_{\rm olig}$.
Each panel combines integrations performed with a variety of $\sigma$'s
between $0.001$ and $0.1 \gm/{\rm cm}^2$.
Kinks in the distributions for Saturn and Neptune are artifacts
of our assumed disk boundaries at $a = 12.5\AU$ and $45\AU$.
Vertical lines mark the current semi-major axes of solar system giants.
Decreasing $N_{\rm olig}$ shrinks the final orbits for Saturn, Uranus, and
Neptune. Final orbits resembling
those of the current solar system are most easily produced
for $N_{\rm olig}=3$.
}
\label{fig_finalA}
\end{figure}

To improve statistics, we ignore these small artifacts of our disk
boundary conditions and pool together $N_{\rm real} = 438$ realizations,
all of which start with $N_{\rm olig} = 5$ and produce 4-planet systems, but
have a variety of $\sigma$'s between 0.001 and $0.1 \gm/{\rm cm}^2$.
{}From this pool we construct the distribution of final semi-major axes
shown in Fig.~\ref{fig_finalA}a.
Clearly, most realizations end with Saturn, Uranus, and Neptune on orbits
too large compared to their actual counterparts in the solar system.
Furthermore, Jupiter typically migrates inward as a consequence
of ejecting several oligarchs outward. The excessively
large orbits of Saturn, Uranus, and Neptune are a consequence
of those planets having scattered oligarchs inward to Jupiter.

\placefigure{fig_finalA_sigma}

\placefigure{fig_finalA}

\begin{figure}
\epsscale{0.9}
%\plotone{RvsLogT_5_0.02_493.eps}
\plotone{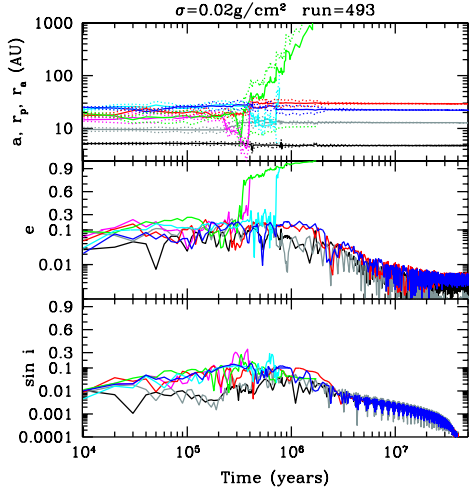}
\caption{Example realization for $\sigma = 0.02\gm/{\rm cm}^2$
and $N_{\rm olig}=5$ in which 3 oligarchs (denoted by
magenta, green, and cyan curves) escape before
the system stabilizes. The remaining 2 oligarchs (blue and red) have final,
nearly circular and co-planar orbits inside 30 AU, similar to those of
Uranus and Neptune. Such an outcome occurs with a frequency of a few
percent for $\sigma = 0.02\gm/{\rm cm}^2$.
}
\label{fig_sample_3}
\end{figure}

Though they only comprise (given our assumed initial conditions)
a few percent of outcomes for $\sigma = 0.02 \gm/{\rm cm}^2$,
some realizations better resemble the solar system
insofar as Uranus and Neptune have final semi-major axes less than 30 AU.
Fig.~\ref{fig_sample_3} showcases an example. Even for this simulation,
however, Saturn's final orbit is 3 AU larger than its actual one.

\placefigure{fig_sample_3}

Reducing the number of starting oligarchs significantly lessens the problem
of excessive migration. Figs.~\ref{fig_finalA}b and \ref{fig_finalA}c
show final semi-major axis distributions corresponding to $N_{\rm olig} = 4$
and 3, respectively. Outcomes for $N_{\rm olig} = 3$
are most solar-system-like. In \S\ref{sec_compact} we will experiment
with initial conditions to demonstrate that a level of agreement
comparable to that displayed in Fig.~\ref{fig_finalA}c for $N_{\rm olig}=3$
can also be obtained for $N_{\rm olig}=4$.

\subsubsection{Stirring of KBOs}
\label{sec_stir}

\begin{figure}
%\plotone{KBOs_5_0.02_493.eps}
\plotone{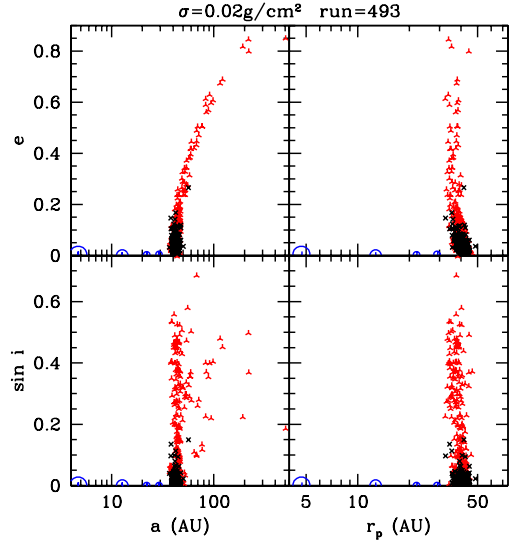}
\caption{Orbital elements of simulated KBOs (black crosses)
at $t = 5 \times 10^7\yr$, for
the same simulation shown in Fig.~\ref{fig_sample_3}. Eccentricities
and inclinations are displayed against either semi-major axis $a$
or perihelion distance $r_{\rm p}$. For comparison we show orbital
elements of known classical and scattered (non-resonant) KBOs
(red inverted Y's) from C06. Orbital elements of surviving planets
(Jupiter, Saturn, and the two oligarchs) are shown as blue circles.
Simulated KBOs for this run
do not exhibit the large eccentricities and inclinations
characterizing actual KBOs, but see Figs.~\ref{fig_KBOs_best} and
\ref{fig_KBOs_compact} for other examples.
}
\label{fig_KBOs_3}
\end{figure}

Fig.~\ref{fig_KBOs_3} describes how the test particles (KBOs),
initially distributed in a dynamically cold ring at $a=40$--45 AU,
are stirred by oligarchs,
for the same $N_{\rm olig}=5$ simulation (Fig.~\ref{fig_sample_3})
which places Uranus and Neptune on final orbits inside 30 AU.
Simulation data are collected at $t = 5 \times 10^7\yr$.
For comparison, Fig.~\ref{fig_KBOs_3}
also plots data for actual KBOs that do not reside
in any strong mean-motion resonance. These objects, taken
from C06, comprise both low-$e$ classical and high-$e$ scattered
KBOs as classified by the Deep Ecliptic Survey (Elliot et al.~2005).
The simulated test particles
have their eccentricities and inclinations
excited up to $\sim$0.1, values that match actual
classical KBOs. But the simulated particles fail to embody the extreme degree
of dynamical heating exhibited by scattered KBOs. Marauding
oligarchs in this simulation stir planetesimals at 40--45 AU too briefly.

\placefigure{fig_KBOs_3}

\begin{figure}
%\plotone{RvsLogT_3_0.04_109.eps}
\plotone{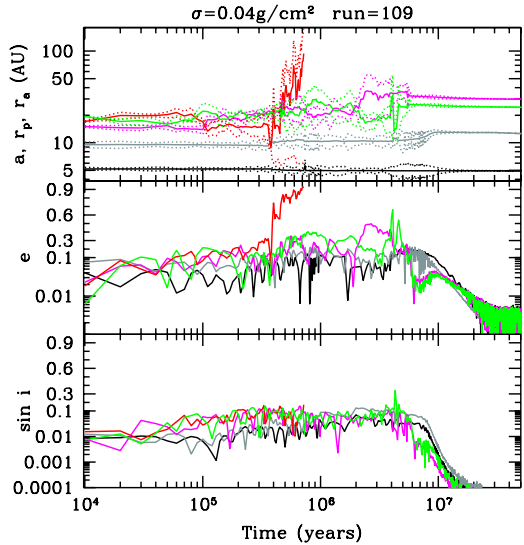}
\caption{Example realization for $N_{\rm olig}=3$ and
$\sigma = 0.04\gm/{\rm cm}^2$ in which surviving
planets have final orbits resembling those of the solar system (for an even
closer match, see Fig.~\ref{fig_RvsT_compact}, generated using a revised set
of initial conditions).
Moreover, the KBOs in this simulation, shown in Fig.~\ref{fig_KBOs_best},
are stirred considerably by oligarchs. Such an outcome
represents several percent of runs generated using $N_{\rm olig}=3$.
}
\label{fig_RvsT_best}
\end{figure}

\begin{figure}
%\plotone{KBOs_3_0.04_109.eps}
\plotone{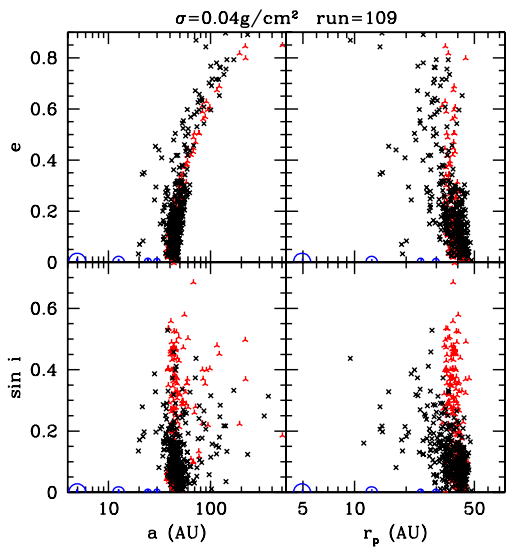}
\caption{Orbital elements of simulated KBOs at $t = 5 \times 10^7\yr$, for
the same simulation shown in Fig.~\ref{fig_RvsT_best}.
In every panel except the one at the bottom left, simulated KBOs (black
crosses)
located to the left of observed KBOs
(red inverted Y's) are unstable over the age of the solar system because
of perturbations by the giant planets. Stable simulated KBOs (lying on top
of and to the right of observed KBOs) exhibit large eccentricities
and inclinations as a consequence of scattering by velocity-unstable oligarchs.
Nevertheless, the simulations fail to reproduce the most extreme of observed
scattered KBOs having $\sin i \gtrsim 0.2$.
}
\label{fig_KBOs_best}
\end{figure}

Since simulations with $N_{\rm olig}=3$ more efficiently
generate solar-system-like planetary spacings than do simulations with
$N_{\rm olig}=5$, we can more thoroughly map out the possible extents to which
KBOs are stirred for $N_{\rm olig}=3$. Figs.~\ref{fig_RvsT_best}
and \ref{fig_KBOs_best} document one simulation, representative
of several percent of the solar-system-like realizations generated
using $N_{\rm olig}=3$, in which KBOs are stirred considerably.
Even here, however, the proportion of simulated KBOs
that simultaneously attain inclinations $i \gtrsim 10^{\circ}$
and perihelion distances $r_{\rm p} \gtrsim 35$ AU is less than observed.
The proportion of simulated KBOs having eccentricities $e \gtrsim 0.3$
also seems under-represented.

\placefigure{fig_RvsT_best}

\placefigure{fig_KBOs_best}

\subsection{More Compact Initial Conditions}
\label{sec_compact}

As described in \S\S\ref{sec_require}--\ref{sec_two},
the simulations that begin with $N_{\rm olig} = 4$ or 5 oligarchs
often do yield 4-planet systems, but the orbital spacings
of the resultant systems do not match those of the solar system.
Multiple ejections displace Jupiter too far inward and displace Saturn,
Uranus, and Neptune too far outward. The problem of excessive
spreading is exacerbated by the need to have Neptune conclude
its orbital evolution by migrating
smoothly and slowly outward from $a\approx 23$ to 30 AU to produce
the population of resonant KBOs (Murray-Clay \& Chiang 2006, and
references therein; but see Levison et al.~2006 for an alternative
theory for the origin of resonant KBOs). This last constraint implies
that our simulations should place Neptune on a final orbit
near $a\approx 23$ AU.

In this sub-section we attempt to remedy the problem of excessive spreading by
adjusting our initial conditions. In anticipation of Jupiter's inward
displacement, we locate that planet initially at $a_{\rm J} = 5.7\AU$.
In anticipation of Saturn's outward displacement, we set $a_{\rm S} = 8\AU$
initially. The innermost oligarch is also shifted inwards, to $a_1 = 12 \AU$.
Initial semi-major axes for remaining oligarchs are still given by
eqn.~(\ref{eqn_define_a}): $a_2$ through $a_5$ equal
13.7, 15.5, 17.7, and 20.1 AU. Finally, so that all oligarchs
lie initially inside the disk, we extend the inner edge of the disk inward
to 10 AU. All remaining parameters remain unchanged from their values
in \S\ref{sec_init}.

\begin{figure}[ht]
%\plotone{FinalACompact.eps}
\plotone{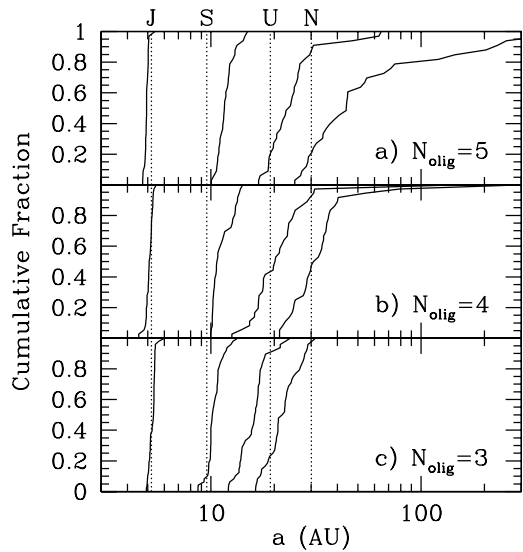}
\caption{Same as Fig.~\ref{fig_finalA}, but generated using the
more compact initial configuration described in \S\ref{sec_compact}.
Runs starting with $N_{\rm olig}=5$ oligarchs still produce final orbits
for the 2 surviving oligarchs
that are still too large compared to those of ice giants in our solar system.
The case $N_{\rm olig}=3$ produces ice giant
orbits that are too small.
Nevertheless, $N_{\rm olig}=3$ is acceptable if we allow for subsequent
outward migration of the ice giants. Such outward migration
seems necessary to explain the origin of resonant KBOs (C06; but
see Levison et al.~2006 for an alternative theory).
Results for $N_{\rm olig}=4$ are intermediate and produce
orbital spacings most closely resembling those of the current solar system.
Data for $N_{\rm olig}=5$, 4, and 3 are generated
with runs having $\sigma = 0.01$, 0.04, and $0.1 \gm/{\rm cm}^2$, respectively.
}
%N_olig = 5 sigma = 0.01, N_olig=4 sigma=0.04, N_olig=3 sigma=0.1
\label{fig_FinalA_compact}
\end{figure}

The distribution
of final semi-major axes for resultant 4-planet systems is given by
Fig.~\ref{fig_FinalA_compact}, constructed in similar
fashion to Fig.~\ref{fig_finalA}. The more compact initial configuration
produces reasonably close matches to current
orbital spacings in the solar system
for $N_{\rm olig} = 4$. Results for $N_{\rm olig}=3$ are also acceptable
if we allow for the subsequent outward migration of Neptune that is seemingly
demanded by resonant KBOs (see first paragraph of this sub-section).
The case $N_{\rm olig} = 5$ still suffers,
however, from excessive spreading.

\placefigure{fig_FinalA_compact}

\begin{figure}[ht]
%\plotone{RvsLogT_4_0.04_479.eps}
\plotone{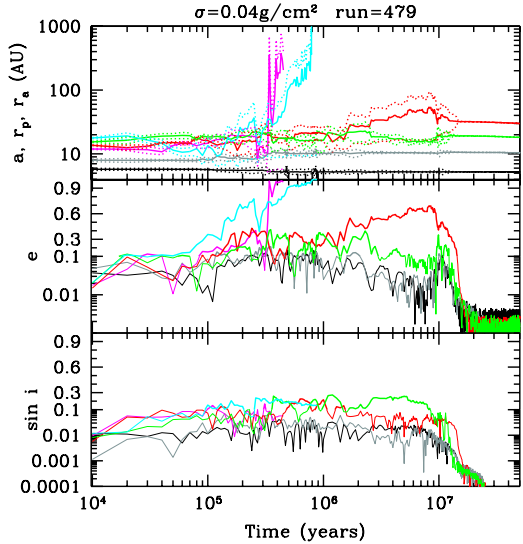}
\caption{Similar to Fig.~\ref{fig_RvsT_best}, except generated
for $N_{\rm olig}=4$ using the more compact initial configuration
described in \S\ref{sec_compact}. The simulation starts with a total
of 6 planets (Jupiter, Saturn, and 4 oligarchs) and ends
with Jupiter, Saturn, and two oligarchs on orbits that closely
match their actual counterparts in the solar system.
}
\label{fig_RvsT_compact}
\end{figure}

\begin{figure}[ht]
%\plotone{KBOs_4_0.04_479.eps}
\plotone{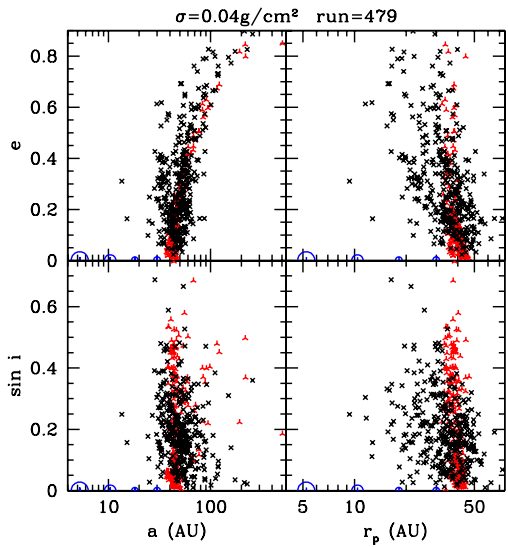}
\caption{Orbital elements of simulated KBOs at $t = 5 \times 10^7\yr$, for
the same simulation shown in Fig.~\ref{fig_RvsT_compact}.
The same remarks
given in the caption to Fig.~\ref{fig_KBOs_best}
concerning the stability of simulated KBOs
apply here. Stable simulated KBOs, though stirred considerably
by velocity-unstable oligarchs, fail to embody
the large inclinations $i \gtrsim 20^{\circ}$ exhibited
by observed scattered KBOs.
}
\label{fig_KBOs_compact}
\end{figure}

Figs.~\ref{fig_RvsT_compact} and \ref{fig_KBOs_compact} sample one
simulation using the revised compact configuration for $N_{\rm olig}=4$.
We highlight this simulation because it reproduces
solar system properties, insofar as (1) Uranus and Neptune have
final semi-major axes less than 30 AU, and (2) the Kuiper belt at
40--45 AU is significantly stirred. Though outcome (1) is
not infrequent---occurring in, e.g., 16 out of 100 runs with
$\sigma = 0.04\gm/{\rm cm}^2$ and $N_{\rm olig} = 4$---outcome (2) is
less probable, characterizing only several percent of runs already
culled to satisfy (1). Most runs that satisfy (1) stir KBOs to
eccentricities and inclinations of just a few percent.
By contrast, the simulation showcased in
Fig.~\ref{fig_KBOs_compact} excites large eccentricities
and inclinations similar to those sported by actual KBOs.
Nevertheless, the most extreme of scattered KBOs,
having perihelion distances $r_{\rm p} \gtrsim 40$ AU and
inclinations $i > 20^{\circ}$, are still under-represented.
In short, our revised compact configuration
stirs KBOs to about the same degree as our original configuration.

\placefigure{fig_RvsT_compact}

\placefigure{fig_KBOs_compact}

\section{SUMMARY AND OUTLOOK}
\label{sec_sum}

In \S\ref{sec_answers}, we answer the three questions
posed in \S\ref{sec_intro}. In \S\ref{sec_commentary},
we place our work in a broader context and mention some directions
for future work.

\subsection{Answers to Questions Posed in \S\ref{sec_intro}}
\label{sec_answers}

\begin{enumerate}

\item Of all our simulations that initially place
$N_{\rm olig}=5$ Neptune-mass planets between 15 and 25 AU,
and that have disk surface densities
$\sigma \approx 0.1 \gm/{\rm cm}^2 \approx 0.07 \Sigma$ (where
$\Sigma \sim 1.5 \gm/{\rm cm}^2$ is the original surface density in oligarchs),
50\% result in the ejection of a single oligarch (Fig.~\ref{fig_branching}).
For runs that begin with
$N_{\rm olig}=4$ and 3 oligarchs, we achieve similar outcomes for
$\sigma/\Sigma \sim 0.05$ and $\sim$$0.03$, respectively.
Jupiter is responsible for the vast majority of ejections,
which occur within $\sim$$10^7\yr$.
The likelihood of a single ejection remains as high as 20\% if the
above $\sigma$'s are increased by factors of 2--3.
Roughly speaking then, we find that instability and ejection require
$\sigma/\Sigma \lesssim 0.1$. By comparison,
GLS04 estimate that $\sigma/\Sigma \lesssim 1$ for instability.
The difference arises partly
because nearest-neighboring oligarchs in our simulations are
separated by 5 Hill sphere radii, whereas
their analysis of shear-dominated oligarchy
assumes the separation is closer to $\sim$1 Hill sphere radius.
Our choice of $5 R_{\rm H}$ is motivated by the half-width
of an oligarch's annular feeding zone in a shear-dominated disk. This
half-width spans $2.5R_{\rm H}$ (Greenberg et al.~1991).
Because oligarchs separated
by $5R_{\rm H}$ viscously stir each other more slowly (Fig.~\ref{fig_stir},
phase 2)
than do oligarchs separated by $1R_{\rm H}$, we find
a threshold value for $\sigma$ lower than what GLS04 estimate.

\item For certain choices of $\sigma$ and initial
semi-major axes, systems starting with $N_{\rm olig}=3$ or 4 oligarchs
frequently end with 2 surviving oligarchs on nearly circular
and co-planar orbits inside 30 AU (Figs.~\ref{fig_finalA}
and \ref{fig_FinalA_compact}). For example, of all runs
that (a) use our revised set of initial semi-major axes
(\S\ref{sec_compact}), (b) begin with $N_{\rm olig}=4$ oligarchs,
and (c) have $\sigma = 0.04\gm/{\rm cm}^2 \approx 0.02 \Sigma$,
44\% end with solar-system-like configurations in which the outermost
surviving oligarch orbits inside 30 AU. This percentage decreases
with increasing $N_{\rm olig}$. This is because
surviving oligarchs spread outward, well beyond
the current orbit of Neptune, as they scatter
more oligarchs inward for eventual ejection by Jupiter.
To eject efficiently more than one oligarch requires that
$\sigma$ be reduced considerably below the previously
mentioned threshold of $\sim$$0.1\Sigma$.
For example, we find that for
$N_{\rm olig}=4$ and our original set of initial semi-major axes
(\S\ref{sec_init_con}),
setting $\sigma \sim 0.02 \gm/{\rm cm}^2 \sim 0.01\Sigma$
maximizes the likelihood of 2 ejections at $\sim$50\%.

\item In a small fraction of
runs that successfully place Jupiter, Saturn, and 2 oligarchs
on solar-system-like orbits inside 30 AU, test particles (KBOs)
located initially in a dynamically cold ring at 40--45 AU have
their eccentricities and inclinations considerably excited by
velocity-unstable oligarchs. We observe maximum eccentricities
of $\sim$0.8 and maximum inclinations of $\sim$$20^{\circ}$
(Figs.~\ref{fig_KBOs_best} and \ref{fig_KBOs_compact}).
In runs characterized by the greatest degrees of excitation,
orbits of simulated KBOs resemble those of observed
classical KBOs and some observed scattered KBOs.
However, no run reproduces the large proportion of observed scattered KBOs
having inclinations $\gtrsim 20^{\circ}$. There may also be
a problem in generating enough KBOs having
eccentricities $\gtrsim 0.3$, given the observational
selection bias against finding such objects.

\end{enumerate}

These results are complementary to those of Levison \& Morbidelli (2007),
who concentrate on the limit $\sigma \gg \Sigma$ and find that
they cannot produce solar-system-like outcomes.
In comparison, we study the case $\sigma \lesssim \Sigma$
and find positive results.

\subsection{Commentary}
\label{sec_commentary}

\subsubsection{The Compactness of Our Preferred Initial Conditions}
We find that
a shear-dominated oligarchy can readily produce a solar-system-like outcome
if it contains just a few excess oligarchs---about 1 or 2 extra---and
if the oligarchs initially reside inside 20 AU.
We are driven to these parameters because to scatter excess oligarchs
inward (toward Jupiter for eventual ejection), surviving oligarchs must be
scattered outward. The right amount of outward spreading is achieved
for suitably compact initial configurations and not too many ejections.

Our favored initial conditions are about as compact as those of
Tsiganis et al.~(2005), who place their outermost ice giant
initially at 17 AU. By comparison, in our revised set
of initial conditions for $N_{\rm olig}=4$, the outermost
oligarch is located at 17.7 AU. But we stress that our study
differs from theirs in that we base our initial conditions on considerations of
shear-dominated oligarchic accretion. An ice giant cannot
form at 17 AU within the gas photoevaporation time of a few $\times$ $10^7\yr$
without strong gravitational focussing (\S\ref{sec_intro}; Levison \& Stewart
2001; GLS04). This need for gravitational focussing can be met by a
highly dissipative disk of planetesimals (GLS04; Rafikov 2004).
It is this disk, and the multiple ($>2$) ice giants that it spawns,
that we have modeled.

Why such a disk would not
form Neptune-mass oligarchs outside 20 AU is an open question.
For some ideas on what limits the sizes of planetary systems,
see Youdin \& Shu (2002) and Youdin \& Chiang (2004).

\subsubsection{Reducing the Disk Surface Density (``Clean-Up'' and Migration)}
\label{sec_reduce}

Velocity instability and the ejection of a single oligarch
require $\sigma/\Sigma \lesssim 0.1$. Ejecting more than 1 oligarch
requires still lower values of $\sigma/\Sigma \sim 0.01$. How can $\sigma$
reach
such low values? Accretion and/or ejection of planetesimals by oligarchs
are natural possibilities. The rate at which $\sigma$ decreases,
compared to the rate at which oligarchs stir each other,
determines whether more than 1 oligarch escapes. If the rate
of depletion of $\sigma$ is sufficiently slow, then after the first oligarch
escapes, surviving planets may occupy orbits so spread apart
that they remain stable even as $\sigma$ decreases further.
On the other hand,
if the rate of depletion is fast, then conditions required to eject more
than 1 oligarch can be met.
We leave investigation of the time dependence
of $\sigma$ for future work.

Given an initial surface density in oligarchs of $\Sigma \sim 1 \gm/{\rm
cm}^2$,
the disk surface densities relevant for instability and ejection range
from $\sigma \approx 0.1$ to $0.01 \gm/{\rm cm}^2$.
These $\sigma$'s
are still higher than surface densities characterizing the Kuiper
belt today, which are on the order of $0.001 \gm/{\rm cm}^2$ (integrated
over all KBO sizes).
``Cleaning up'' the disk mass remains an unsolved problem (GLS04).
Again, the mass could either be accreted or ejected by surviving planets.
Planetesimal ejection drives planetary migration. For Neptune to expand
its orbit from $\sim$23 to 30 AU, as seemingly demanded by the large
observed population of resonant KBOs (C06), the planet must
scatter at least $\sim$$7/30 \sim 25$\% of its own mass in planetesimals,
or about $4 M_{\earth}$. Spreading such a mass over a disk of radius
30 AU yields a surface density on the order of $0.03 \gm/{\rm cm}^2$,
which falls within the range of $\sigma$'s that we find characterize
instability. Clean-up and migration go hand-in-hand.

The disk masses that we found to be relevant for instability
are smaller than the masses in the planets. This raises concern
about the validity of our approximation that disk properties remain fixed
throughout the simulation.
One way of testing this approximation is to compare the change in
the total angular momentum of all planets (brought about by dynamical
friction) to the angular momentum
available in the disk. If the former were much larger than the latter,
then our neglect of back-reaction upon the disk would be a poor assumption.
For the simulations displayed in Figs.~\ref{fig_RvsT_best}
and \ref{fig_RvsT_compact}, we find
that the change in the $z$-component of angular momentum of all
planets (including ejected ones) is nearly identical to the angular
momentum available in the disk, indicating that our assumption
of a fixed disk may be only marginally valid. (An analogous test for
the energy would be inconclusive since the disk is supposed to be
a sink of energy by virtue of dissipative collisions).
For ideas on how to treat planet-disk interactions self-consistently,
see Lithwick \& Chiang (2007) and Levison \& Morbidelli (2007).

\subsubsection{Disk Optical Depths and Comparison to Debris Disks}
A planetesimal disk of
surface density $\sigma$ has a geometric vertical optical depth
$\tau_p \sim 4 \times 10^{-6} [\sigma/(0.01 \gm/{\rm cm}^{2})] (10 \m / p)$,
where $p$ is the assumed planetesimal radius.
Collisions between planetesimals,
which occur over a timescale $\sim$$1/(\Omega\tau_p) \sim 10^7 \yr$ at 30 AU,
generate smaller dust particles whose optical
depth is orders of magnitude higher. For example,
if the dust size distribution obeys a Dohnanyi (1969) spectrum,
then the geometric, vertical optical depth in $s$-sized grains would be
$\tau_d \sim \tau_p (p/s)^{1/2} \sim 0.01 [\sigma/(0.01 \gm/{\rm cm}^2)] (10\m
/ p)^{1/2} (\mu{\rm m}/s)^{1/2}$.
This is comparable to the vertical optical depths of some of the brightest
extra-solar debris disks observed, e.g., $\beta$ Pic (Artymowicz 1997),
HR 4796A (Li \& Lunine 2003), and AU Mic (Strubbe \& Chiang 2006),
systems that are all $\sim$$10^7\yr$ old.
The observed paucity of stars with optically
thicker disks implies that large populations of planetesimals having
sizes $p < 10 \m$ at stellocentric distances of $\sim$30 AU cannot be
maintained for longer than $\sim$$10^7\yr$. The surface density in
such collisional objects must be reduced by at least 2 orders of
magnitude below planet-forming values of $\sim$$1 \gm/{\rm cm}^2$
within this timescale. In other words, planetesimals that are both
collisional and planet-forming, like the kind espoused by GLS04, must be
cleaned up fairly quickly.

\subsubsection{Instability in Our Solar System and Others}
\label{sec_instab_ss}

We have demonstrated that more than 2 ice giants
may once have orbited the Sun.
The current architecture of the outer solar system may well have
resulted from a prior era of dynamical instability
during which Uranus, Neptune, and 1 or 2 other
ice giants crossed paths.

We expect similar instabilities to afflict all nascent planetary systems.
Perhaps planet-planet instabilities are reflected in the
large orbital eccentricities exhibited by
most extra-solar gas giants (Marzari \& Weidenschilling 2002; Ford et al.~2003;
but for an alternative view see Goldreich \& Sari 2003).
The case of Upsilon Andromedae fits this picture (Ford et al.~2005).
The difference between our solar system and systems like Ups And
might be the surface density of the parent disk at the time
of the last planet-planet scattering (Ford 2006).
The time of last scattering will vary widely
because of the chaotic nature of multi-planet systems.
In the case of the solar system, the disk surface density
must have been large enough at the time of last scattering
for dynamical friction to damp
the eccentricities and inclinations of surviving planets
back down.

Viscous stirring rates vary with the semi-major axis
separation between oligarchs.
At least in the case without gas giants, the
time for oligarchs to undergo close encounters
increases by several orders of magnitude as
their semi-major axis spacing is increased
from 3 to 7 mutual Hill radii (Chambers et al.~1996).
The disk surface density required for instability
depends directly on this time, i.e.,
$\sigma_{\rm crit} \propto 1/t_{\rm unstable}$
in our eqn.~(\ref{eqn_semi_empirical}).
Whether $t_{\rm unstable}$ varies as strongly with oligarch
spacing when perturbations by Jupiter and Saturn are included
is not known, but preliminary experiments by us suggest
that it does not. When we change the oligarch spacing
from our standard 5 Hill radii to 3 Hill radii in runs that include
Jupiter, Saturn, and $N_{\rm olig}=5$ oligarchs,
we find that the probability of 1 ejection still
peaks at $\sim$50\% for $\sigma/\Sigma \approx 0.1$ (accounting for
the factor of 2 increase in $\Sigma$ due to the shorter spacing).

\subsubsection{Evidence for the Velocity Instability in the Kuiper Belt}
Did velocity-unstable ice giants excite
the large eccentricities and inclinations of
the scattered Kuiper belt, as proposed by C06?
Our provisional answer is no, as we were unable to reproduce
the large inclinations of scattered KBOs.
For runs with disk surface densities down to $\sim$$0.01 \gm /{\rm cm}^2$,
oligarchs spend too little time, less than $\sim$$10^7\yr$,
passing through the Kuiper belt. Moreover,
oligarchs in our simulations
have orbital inclinations that rarely exceed $10^{\circ}$.

To remedy the situation, we might appeal to still lower disk surface
densities, on the order of $\sim$$0.001 \gm/{\rm cm}^2$,
for which dynamical friction cooling times for embedded planets
would be as long as $\sim$$10^8\yr$. We found this region of
parameter space difficult to explore. Out of 100 simulations
starting with (a) $N_{\rm olig}=4$, (b) our compact set of initial conditions,
and (c) $\sigma = 0.002 \gm/{\rm cm}^2$, only 6 yielded systems each
with 2 surviving oligarchs at the end of the integrations at $t = 10^8\yr$.
Unfortunately, 5 of these 6 systems had not stabilized, and it was
unclear whether more oligarchs would be ejected were the integrations
to continue. Furthermore, over these long timescales,
effects resulting from time variations in $\sigma$
(see \S\ref{sec_reduce}) might be expected to be important,
and we have not modeled these.

Cooling times for planets and, by extension,
KBO heating times might also be prolonged in
more realistic treatments of dynamical friction
that incorporate non-axisymmetries and the clearing of gaps
in the disk.
Ways of numerically simulating the response of planetesimal
disks to planets can be found in Levison \& Morbidelli (2007)
and Lithwick \& Chiang (2007).

\acknowledgments
We thank Ben Collins, Yoram Lithwick, Ruth Murray-Clay,
Re'em Sari, and an anonymous referee for helpful exchanges.
We also thank Harold
Levison and Alessandro Morbidelli for generously sharing some of their
ideas about planet-disk interactions and suggesting that we examine
the angular momentum and energy budgets of our simulations.
Support for E.B.F.\ was provided by a Miller Research Fellowship
and by NASA through Hubble Fellowship grant
HST-HF-01195.01A awarded by the Space Telescope Science Institute,
which is operated by the Association of Universities for Research in
Astronomy, Inc., for NASA, under contract NAS 5-26555.
E.C. acknowledges grants from the National Science
Foundation (AST-0507805), NASA (JPL-1264475), and the Alfred P.~Sloan
Foundation.

\end{document}